\documentclass[journal]{IEEEtran}

\usepackage{xspace,amsmath,amssymb,amsfonts,epsfig,syntonly}
\usepackage{cite,bm,color,url,textcomp,empheq,boxedminipage}
\usepackage{algorithmicx,algorithm,algpseudocode}
\usepackage{epstopdf}
\usepackage{empheq}
\usepackage{graphicx,graphics}  
\usepackage{multirow,multicol}
\usepackage{psfrag}    
\usepackage{stfloats}

\newtheorem{lemma}{\underline{Lemma}}[section]

\newtheorem{remark}{\underline{Remark}}[section]

\newcommand{\mv}[1]{\mbox{\boldmath{$ #1 $}}}

\DeclareMathOperator*{\argmax}{arg\,max}

\long\def\symbolfootnote[#1]#2{\begingroup
\def\thefootnote{\fnsymbol{footnote}}
\footnote[#1]{#2}\endgroup}

\psfull

\usepackage[T1]{fontenc}
\usepackage{ifpdf}
 \ifpdf
 \else
 \fi

\begin{document}
\title{Computation Rate Maximization for Wireless Powered Mobile Edge Computing}
\author{Feng Wang\\
 School of Information Engineering, Guangdong University of Technology, Guangzhou, China\\
  Email: fengwang13@gdut.edu.cn 
\thanks{$^*$The author would like to thank his colleague Prof. Jie Xu for the careful discussions and comments in this work.}}
\maketitle

\begin{abstract}
 Integrating mobile edge computing (MEC) and wireless power transfer (WPT) has been regarded as a promising technique to improve computation capabilities for self-sustainable Internet of Things (IoT) devices. This paper investigates a wireless powered multiuser MEC system, where a multi-antenna access point (AP) (integrated with an MEC server) broadcasts wireless power to charge multiple users for mobile computing. We consider a time-division multiple access (TDMA) protocol for multiuser computation offloading. Under this setup, we aim to maximize the weighted sum of the computation rates (in terms of the number of computation bits) across all the users, by jointly optimizing the energy transmit beamformer at the AP, the task partition for the users (for local computing and offloading, respectively), and the time allocation among the users. We derive the optimal solution in a semi-closed form via convex optimization techniques. Numerical results show the merit of the proposed design over alternative benchmark schemes.
\end{abstract}

\section{Introduction}
As an enabling technique to provide cloud-like computing for various low-latency and computation-extensive Internet of Things (IoT) applications such as augmented reality and autonomous driving, mobile edge computing (MEC) has received growing attentions from both industry and academia\cite{Chiang16,Mao17,Feng17_2,Mach17,Bar14,Liu16,Chen16,ETSI,Feng17,Cao17,X_Chen16,You16}. At the edge of radio access networks such as access points (APs), MEC servers are deployed therein. The IoT devices are generally of small size and low power. Depending on the computation task is partitionable or not, the resource-limited IoT devices can offload part or all of their computation tasks to the APs, respectively; then the installed MEC servers can execute the offloaded tasks on behalf of these devices. 

On the other hand, radio-frequency (RF) signal based wireless power transfer (WPT) provides a viable solution for powering self-sustainable IoT devices by deploying dedicated energy transmitters to broadcast energy wirelessly\cite{Xu14,Li15,Feng15}. Wireless powered communication networks (WPCNs) and simultaneous wireless information and power transfer (SWIPT) paradigms can achieve ubiquitous wireless communications in a self-sustainable way. It is also expected that WPT can facilitate self-sustainable ubiquitous computing for IoT devices.

Note that the prior works in \cite{You16,Feng17} investigated the wireless powered MEC systems with one or more wireless IoT devices in a self-sustainable way. Specifically, in \cite{You16} the single user maximizes the probability of successful computation with binary offloading, where each task is not partitionable but offloaded as a whole or locally computed by user itself. Assuming that each task is partitionable and a time-division multiple access (TDMA) protocol is adopted for the partial offloading, \cite{Feng17} minimized the total energy consumption for the AP by jointly optimizing the energy transmit beamforming, the task partition and the central processing unit (CPU) frequency per user, as well as the TDMA based time allocations for computation offloading across the users.

Different from the above works, this paper considers a wireless powered MEC system with {\em limited} resources. The AP employs energy transmit beamforming to simultaneously charge multiple users for mobile computing. The downlink WPT and the computation offloading are operated over orthogonal frequency bands. Suppose that the partial offloading is allowed for each user. As in \cite{Feng17}, a TDMA protocol is employed to coordinate multiuser computation offloading. Under this setup, we aim to maximize the weighted sum of computation rates (in terms of the number of computation bits over a particular time block) across all the users subject to the limited MEC computation capacity and the AP transmit power budget constraints. We jointly optimize the energy transmit beamformer at the AP, the task partition for local computing and offloading for each user, as well as the time allocation among the users. Using the Lagrange dual method, we derive the optimal solution in a semi-closed form. Numerical results show the merit of the proposed joint design over alternative benchmark schemes.

The remainder of the paper is organized as follows. Section~\ref{Sec:System} presents the system model and formulates the weighted sum of computation rates maximization problem of our interest. Section~\ref{Sec:Optimal} develops an efficient algorithm to obtain the optimal solution in a semi-closed form. Section~\ref{Sec:Numerical} provides numerical results, followed by the conclusion in Section~\ref{Sec:Conclusion}.


\section{System Model and Problem Formulation}\label{Sec:System}
Consider a wireless powered multiuser MEC system, where an $N$-antenna AP (with an integrated MEC server) employs RF signal based transmit energy beamforming to charge a set ${\cal K}\triangleq \{1,\ldots,K\}$ of single-antenna users. The downlink WPT and the wireless communication (for offloading) are assumed to operate over orthogonal frequency bands simultaneously. We consider a time block of finite duration $T$ for both the WPT and the MEC. Assume that the task is partitionable for each of the $K$ users. Relying on the harvested energy in one block, each user can then arbitrarily partition its task into two parts for local computing and for offloading to the MEC server, respectively. Note that the computation task at each user must be accomplished before the end of this block; hence, the number of computation bits over the block can measure the computation rate. In addition, we assume that the AP perfectly knows the computation information of all the $K$ users, as well as the channel state information (CSI) from/to the $K$ users. 

\subsection{Energy Transmit Beamforming from AP to Users}
Let $\boldsymbol{s}\in\mathbb{C}^{N\times 1}$ and $\boldsymbol{Q}\triangleq \mathbb{E}[\boldsymbol{s}\boldsymbol{s}^H]$ denote the energy-bearing transmit signal by the AP and the energy transmit covariance matrix, respectively, where $\mathbb{E}[\cdot]$ and the superscript $H$ denote the stochastic expectation and the Hermitian transpose, respectively. Let $P_{\max}$ denote the maximum transmit power at the AP. We then have the following energy transmit beamforming constraint at the AP:
\begin{align}\label{eqn:Q}
{\rm tr}({\mv Q})\triangleq \mathbb{E}[\|{\bm s}\|^2] \leq T P_{\max},
\end{align}
where $\|\cdot\|$ and ${\rm tr}(\cdot)$ denote the Euclidean norm and the trace operation, respectively. In general, the AP can use multiple energy beams to deliver the wireless energy, i.e., $\boldsymbol{Q}$ can be of any rank~\cite{Xu14}. 
Let $\boldsymbol{h}_i\in\mathbb{C}^{N\times 1}$ denote the channel vector from the AP to user $i\in {\cal K}$, and define $\boldsymbol{H}_i\triangleq \boldsymbol{h}_i\boldsymbol{h}_i^H$, $\forall i\in{\cal K}$. The harvested energy amount by user~$i$ over this block is then
\begin{align}\label{eq.energy_harvested}
E_i =  T\eta \mathbb{E}\left[\left| \boldsymbol{h}^H_i\boldsymbol{s} \right|^2\right] = T\eta{\rm tr}(\boldsymbol{Q}\boldsymbol{H}_i),
\end{align}
where $|\cdot|$ denotes the absolute value of a scalar and $0<\eta\leq 1$ is the energy conversion efficiency per user. Both the local computing and offloading for user $i\in{\cal K}$ in the block are powered the harvested energy $E_i$.

\subsection{Computation Task Execution for Users}
Over the duration-$T$ block, the computation task for user $i\in{\cal K}$ is partitioned into two parts with $\ell_i\geq 0$ and $q_i\geq 0$ bits for offloading and local computing, respectively, in parallel.

\subsubsection{Computation Offloading from Users to the AP}
Consider a TDMA protocol for the $K$ users' offloading, where the block is divided into $2K$ time slots each with duration $t_i$, $\forall i\in\{1,\ldots,2K\}$. These users offload their computation bits to the AP one by one in the first $K$ time slots. After the offloaded tasks having been executed at the MEC server, the AP sends the computation results back to the $K$ users in the next $K$ time slots sequentially. As in \cite{Feng17}, the computation time consumed at the MEC server is negligible and the user can download the computation results immediately after the first $K$ time slots, i.e., $t_i\approx 0$, $\forall i\in\{K+1,\ldots,2K\}$. In addition, we ignore the energy consumption for transmitting/receiving the computation results in this paper. As a result, the TDMA based offloading time allocation across the $K$ users is
\begin{align}\label{eqn:T}
\sum_{i=1}^K t_i\leq T.
\end{align}

Let $\boldsymbol{g}_i\in \mathbb{C}^{N\times 1}$ denote the channel vector from user $i$ to the AP and $p_i\geq 0$ the transmit power for user $i$'s offloading in time slot $t_i$. The maximum ratio combining (MRC) receiver is further assumed for the AP to decode the information. The achievable offloading rate (in bits/sec) for user $i$ is then
\begin{align}\label{eq.rate}
r_i = B \log_2\left(1+\frac{p_i\tilde{g}_i}{\Gamma\sigma^2}\right),~~\forall i\in{\cal K},
\end{align}
where $B$ denotes the bandwidth, $\tilde{g}_i \triangleq \|\boldsymbol{g}_i\|^2$ denotes the effective channel power gain from user~$i$ to the AP, $\sigma^2$ is the noise power at the receiver of the AP, and $\Gamma\geq 1$ is a constant accounting for the gap from the channel capacity due to a practical coding and modulation scheme. For simplicity, $\Gamma=1$ is assumed throughout this paper. As a result, the number of offloaded bits $\ell_i$ by user $i$ to the AP can be expressed as
\begin{align}\label{eq.ell}
\ell_i = r_i t_i,~~\forall i\in {\cal K}.
\end{align}
Consider an MEC server with limit computation capacity. Let $L_{\max}$ be the maximum number of computation bits that can be executed at the MEC server over the block. We then have the following computation capacity constraint:
\begin{align}\label{eqn:L}
\sum_{i=1}^K \ell_i \leq L_{\max}.
\end{align}

It is worth noting that computation offloading incurs energy consumption at both the $K$ users and the AP. Per user $i\in{\cal K}$, in addition to the transmit power $p_i$, a constant circuit power $p_{{\rm c},i}> 0$ (by the digital-to-analog converter (DAC), filter, etc.) is consumed. The offloading energy consumption at user $i$ is
\begin{align}\label{eq.energy_offl}
E_{\text{offl},i} = \frac{t_i}{\tilde{g}_i}\beta\left(\frac{\ell_i}{t_i}\right) +p_{c,i} t_i,
\end{align}
where $\beta(x) \triangleq \sigma^2(2^{\frac{x}{B}}-1)$ is a monotonically increasing and convex function with respect to $x$. Note that to avoid the issue of dividing by zero, we define $\beta\left(\frac{\ell_i}{t_i}\right)=0$ when either $\ell_i=0$ or $t_i=0$ holds. 


\subsubsection{Local Computing at Users}
We next consider the local computing for executing $q_i$ computation bits at each user $i\in {\cal K}$. Let $C_i$ be the number of CPU cycles required for computing one computation bit at user $i$. Then the total number of CPU cycles for the $q_i$ bits is $C_iq_i$. By applying the dynamic voltage and frequency scaling (DVFS) technique\cite{Mach17}, user $i$ can adjust the CPU frequency $f_{i,n}$ for each cycle $n\in\{1,\ldots,C_iq_i\}$, where $0<f_{i,n}\leq f_i^{\max}$ and $f_i^{\max}$ denotes user $i$'s maximum CPU frequency. As the local computing should be accomplished before the end of the block, we have the following computation latency requirements:
\begin{align}\label{eq.f_in}
\sum_{n=1}^{C_iq_i} \frac{1}{f_{i,n}}\leq T,~~\forall i\in {\cal K}.
\end{align}
Accordingly, the consumed energy for local computing at user $i\in{\cal K}$ is given by
\begin{align}\label{eq.energy_loc}
E_{{\rm loc},i} =\sum_{n=1}^{C_iq_i} \zeta_i f^2_{i,n},
\end{align}
where $\zeta_i>0$ is the effective capacitance coefficient that depends on the chip architecture at user $i$. In order to minimize the energy consumption while satisfying the latency constraint, it is optimal for each user to set the CPU frequencies to be identical for different CPU cycles (see \cite[Lemma 3.1]{Feng17}). By using this fact and letting the constraints in \eqref{eq.f_in} be met with strict equality, we have
\begin{align}\label{eq.f_i}
f_{i,1}=\cdots=f_{i,C_iq_i}=C_iq_i/T, ~~\forall i\in{\cal K}.
\end{align}
As the maximum CPU frequency $f_i^{\max}$ is specified for each user $i\in{\cal K}$, we have the following constraints for the numbers of computation bits by local computing:
\begin{align}\label{eq.q_i}
0\leq q_i\leq \frac{Tf_{i}^{\max}}{C_i},~~\forall i\in{\cal K}.
\end{align}
By substituting \eqref{eq.f_i} in \eqref{eq.energy_loc}, the energy consumption $E_{{\rm loc},i}$ is re-expressed as
\begin{align}\label{eq.energy_loc2}
E_{{\rm loc},i} =\frac{\zeta_iC_i^3q_i^3}{T^2}.
\end{align}

\subsection{Energy Harvesting Constraints at Users}
 To achieve self-sustainable operation, the energy harvesting constraint needs to be imposed such that the totally consumed energy at the user cannot exceed the harvested energy $E_i$ in (\ref{eq.energy_harvested}) per block\cite{Li15}. By combining the computation offloading energy in (\ref{eq.energy_offl}) and the local computation energy in (\ref{eq.energy_loc2}), the total energy consumed by user $i$ within the block is $E_{\text{offl},i}+E_{\text{loc},i}$. Therefore, we have 
\begin{align}\label{eqn:EH}
\frac{\zeta_iC_i^3 q_i^3}{T^2}+\frac{t_i}{\tilde{g}_i}\beta\left( \frac{\ell_i}{t_i}\right) + p_{c,i}t_i \leq T\eta{\rm tr}\left( \boldsymbol{Q}\boldsymbol{H}_i\right),\forall i\in{\cal K}.
\end{align}

\subsection{Problem Formulation}
We are interested in utilizing the constrained communication/computation resources at the AP to maximize the total computation rates across all the users. Different users usually have different priorities. Accordingly, we maximize the weighted sum-number of users' computation bits subject to the MEC computation capacity and the transmit power constraints at the AP. Let $\omega_i>0$ denote the positive weight for user $i$ that characterizes the priority of its computation task. Mathematically, the weighted sum of computation rates maximization problem is formulated as
\begin{subequations}\label{eq.prob}
\begin{align}
  ({\cal P}1): &\max_{\boldsymbol{Q}\succeq \boldsymbol{0},\boldsymbol{t},\boldsymbol{\ell},\boldsymbol{q}}~ \sum_{i=1}^K \omega_i(q_i+\ell_i) \\
 &{\rm s.t.}~~ 0\leq t_i \leq T, ~~0\leq \ell_i \leq L_{\max},~~\forall i \in {\cal K}\\
 &\quad\quad \eqref{eqn:Q},~\eqref{eqn:T},~\eqref{eqn:L}, ~\eqref{eq.q_i},~{\rm and}~\eqref{eqn:EH}, \notag
\end{align}
\end{subequations}
where $\mv t\triangleq[t_1,\ldots,t_K]^\dagger$, $\mv \ell\triangleq [\ell_1,\ldots,\ell_K]^\dagger$, and $\boldsymbol{q}\triangleq [q_1,\ldots,q_K]^\dagger$ with the superscript $\dagger$ being the transpose operation; $\mv Q\succeq \mv 0$ guarantees $\mv Q$ to be positive semidefinite. Note that problem $({\cal P}1)$ is convex and can thus be efficiently solved by the interior-point method\cite{BoydBook}. Nevertheless, to gain more engineering insights, we next leverage the Lagrange dual method to obtain the optimal solution in a semi-closed form. 

\section{Optimal Solution to Problem $({\cal P}1)$}\label{Sec:Optimal}
In this section, we obtain the optimal solution to $({\cal P}1)$ in a semi-closed form and develop an efficient algorithm.

Let $\rho\geq 0$, $\mu\geq 0$, $\theta\geq 0$, and $\lambda_i\geq 0$ be the Lagrange multipliers for \eqref{eqn:Q}, \eqref{eqn:T}, \eqref{eqn:L}, and the $i$-th constraint in \eqref{eqn:EH}, $\forall i\in{\cal K}$, respectively. The partial Lagrangian of problem $({\cal P}1)$ is given by
\begin{align}\label{eq.par_L2}
& {\cal L}\left(\boldsymbol{Q},\boldsymbol{t},\boldsymbol{\ell},\boldsymbol{q},\boldsymbol{\lambda},\mu,\rho,\theta\right)  = {\rm tr}\left( \left( \sum_{i=1}^K T\eta\lambda_i \boldsymbol{H}_i -\rho\boldsymbol{I}\right) \boldsymbol{Q}\right) \notag \\
 &~+ \mu T +\rho TP_{\max} + \theta L_{\max}+\sum_{i=1}^K \left(  \omega_i q_i -\frac{\lambda_i\zeta_iC_i^3q_i^3}{T^2}\right) \notag \\
&~+\sum_{i=1}^K\left( (\omega_i-\theta)\ell_i-\frac{\lambda_it_i}{\tilde{g}_i}\beta\left(\frac{\ell_i}{t_i}\right)-\mu t_i-\lambda_ip_{{\rm c},i} t_i
 \right).
\end{align}
The dual function of problem $({\cal P}1)$ is then
\begin{align}\label{eq.dual_obj2}
\Phi({\bm \lambda},\mu,\rho,\theta) =& \max_{\boldsymbol{Q}\succeq \boldsymbol{0},\boldsymbol{t},\boldsymbol{\ell},\boldsymbol{q}} ~{\cal L}_2\left(\boldsymbol{Q},\boldsymbol{t},\boldsymbol{\ell},\boldsymbol{q}, \boldsymbol{\lambda},\mu,\rho,\theta \right) \\
{\rm s.t.}& \quad (\ref{eq.q_i})~{\rm and}~(\ref{eq.prob}\rm b). \notag
\end{align}
Consequently, the dual problem of $({\cal P}1)$ is expressed as
\begin{subequations} \label{eq.dual_prob2}
\begin{align}
 ({\cal D}1): ~&\min_{\boldsymbol{\lambda},\mu,\rho,\theta} ~~\Phi({\bm \lambda},\mu,\rho,\theta) \\
 &{\rm s.t.} ~~
 {\bm G}({\bm \lambda},\rho)\triangleq \sum_{i=1}^K \eta\lambda_i \boldsymbol{H}_i - \rho \boldsymbol{I} \preceq \boldsymbol{0}\\
& \quad \quad \mu \geq 0,~ \rho \geq 0,~\theta\geq 0,~ \lambda_i \geq 0 ,~\forall i\in {\cal K}.
\end{align}
\end{subequations}
Note that the constraint (\ref{eq.dual_prob2}b) is to ensure that the dual function $\Phi({\bm \lambda},\mu,\rho,\theta)$ is bounded from above (see Appendix A). Denote by $\cal S$ the set of $(\boldsymbol{\lambda},\mu,\rho,\theta)$ characterized by (\ref{eq.dual_prob2}b) and (\ref{eq.dual_prob2}c).

Since problem $({\cal P}1)$ is convex and satisfies the Slater's condition, strong duality holds between $({\cal P}1)$ and $({\cal D}1)$\cite{BoydBook}. As a result, we can solve $({\cal P}1)$ by equivalently solving the dual problem $({\cal D}1)$. For convenience of presentation, we denote $(\mv Q^*,\mv t^*,\mv \ell^*, \mv q^*)$ as the optimal solution to problem \eqref{eq.dual_obj2} under given $(\boldsymbol{\lambda},\mu,\rho,\theta)\in{\cal S}$, $(\boldsymbol{Q}^{\rm{opt}},\boldsymbol{t}^{\rm{opt}},\boldsymbol{\ell}^{\rm{opt}},\boldsymbol{q}^{\rm{opt}})$ as the optimal solution to $({\cal P}1)$, $(\boldsymbol{\lambda}^{\rm opt},\mu^{\rm opt},\rho^{\rm opt},\theta^{\rm opt})$ as the optimal solution to $({\cal D}1)$.


\subsection{Evaluating the Dual Function $\Phi({\bm \lambda},\mu,\rho,\theta)$}

First, we obtain the dual function $\Phi({\bm \lambda},\mu,\rho,\theta)$ under any given $(\boldsymbol{\lambda},\mu,\rho,\theta)\in{\cal S}$. Note that problem (\ref{eq.dual_obj2}) can be decomposed into $(2K+1)$ subproblems as follows, one for optimizing $\boldsymbol{Q}$, $K$ for optimizing $q_i$'s, and another $K$ for jointly optimizing $t_i$'s and $\ell_i$'s, respectively.
\begin{align}\label{eq.Q2_iter}
\max_{\boldsymbol{Q}}&~~{\rm tr}\left( {\bm G}({\bm \lambda},\rho)\boldsymbol{Q} \right) ~~~~{\rm s.t.} ~~ \boldsymbol{Q}\succeq \boldsymbol{ 0} \\
\max_{q_i}&~~
\omega_iq_i - \frac{\lambda_i \zeta_i C^3_iq_i^3}{T^2} ~~{\rm s.t.} ~~0\leq q_i\leq \frac{Tf_i^{\max}}{C_i} \label{eq.La2_q} \\
\max_{t_i,\;\ell_i}&~~
 (\omega_i-\theta) \ell_i - \frac{\lambda_i t_i}{\tilde{g}_i}\beta\left(\frac{\ell_i}{t_i}\right)-\mu t_i-\lambda_ip_{{\rm c},i}t_i \notag \\
{\rm s.t.} &~~0\leq t_i \leq T,~~0\leq \ell_i\leq L_{\max},\label{eq.La2_t_l}
\end{align}
where $i\in{\cal K}$.

For problem (\ref{eq.Q2_iter}), under the condition of ${\bm G}({\bm \lambda},\rho) \preceq \boldsymbol{0}$, the optimal value of (\ref{eq.Q2_iter}) is zero and the optimal solution $\boldsymbol{Q}^{*}$ to (\ref{eq.Q2_iter}) can be {\em any} positive semidefinite matrix in the null space of ${\bm G}({\bm \lambda},\rho)$. Note that the optimal solution $\boldsymbol{Q}^{*}=\boldsymbol{0}$ of (\ref{eq.Q2_iter}) is used only for evaluating the dual function $\Phi({\bm \lambda},\mu,\rho,\theta)$. 

Next, consider the $i$-th subproblems in (\ref{eq.La2_q}) and (\ref{eq.La2_t_l}). As both problems are convex and satisfy the Slater's condition, one can obtain their solutions in semi-closed forms based on the Karush-Kuhn-Tucker (KKT) conditions~\cite{BoydBook}, as stated in the following two lemmas, respectively.
\begin{lemma}\label{lem.q}
Under given $({\bm \lambda},\mu,\rho,\theta)\in{\cal S}$, the optimal solution of the number of local computing bits $q_i^{*}$ to problem~(\ref{eq.La2_q}) can be obtained as follows.
\begin{itemize}
\item If $\lambda_i=0$, we have $q_i^{*}={Tf_i^{\max}}/{C_i}$;
\item If $\lambda_i>0$, we have
\begin{equation}
q_i^{*} = \min \left( \sqrt{\frac{\omega_i T^2}{3\lambda_i\zeta_iC_i^3}}, \; \frac{Tf_i^{\max}}{C_i} \right).
\end{equation}
\end{itemize}
\end{lemma}
\begin{IEEEproof}
See Appendix B.
\end{IEEEproof}

\begin{lemma}\label{lem.t}
The optimal solution of the offloading time $t_i^{*}$ and the number of offloading bits $\ell_i^{*}$ to problem~(\ref{eq.La2_t_l}) are given below.
\begin{itemize}
\item If $\lambda_i=0$ and $\omega_i-\theta>0$, we have $t_i^{*}=0$ and $\ell_i^{*}=L_{\max}$;
\item If $\lambda_i=0$ and $\omega_i-\theta\leq 0$, we have $t_i^{*}=0$ and $\ell_i^{*}=0$;
\item If $\lambda_i>0$ and $\omega_i-\theta\leq \lambda_i\sigma^2\ln2/(B\tilde{g}_i)$, we have $t_i^{*}=0$ and $\ell_i^{*}=0$;
\item If $\lambda_i>0$ and $\omega_i-\theta> \lambda_i\sigma^2\ln2/(B\tilde{g}_i)$, the optimal offloading rate is given by
    \begin{equation}\label{eq.opt_r2}
    r_i^{*} = B\log_2\left(\frac{(\omega_i-\theta)B\tilde{g}_i}{\lambda_i\sigma^2\ln2}\right),
    \end{equation}
 and $\ell_i^{*}=r_i^{*}t_i^{*}$, where
 \begin{equation}\label{eq.non-unique-t}
t_i^{*}
\left\{ \begin{aligned}
 = &0,&&{\rm if}~\frac{\lambda_i}{\tilde{g}_i}\left(\beta(r^{*}_i)-r^{*}_i\beta'(r^{*}_i)\right)+\mu+\lambda_ip_{{\rm c},i}> 0 \\
 \in &[0,T],&&{\rm if}~\frac{\lambda_i}{\tilde{g}_i}\left(\beta(r^{*}_i)-r^{*}_i\beta'(r^{*}_i)\right)+\mu+\lambda_ip_{{\rm c},i}= 0 \\
=& T,&&{\rm otherwise},
 \end{aligned} \right.
 \end{equation}
 and $\beta'(x)$ denotes the first-order derivative with respect to $x$.
 \end{itemize}
 \end{lemma}

\begin{IEEEproof}
See Appendix C.
\end{IEEEproof}
As stated in Lemma~\ref{lem.t}, if $\frac{\lambda_i}{\tilde{g}_i}\left(\beta(r_i)-r_i\beta'(r_i)\right)+\mu+\lambda_ip_{{\rm c},i} = 0$, $t_i^{*}\in[0,T]$ is generally {\em not} a unique solution; in this case, we set $t_i^{*}=0$ to facilitate the dual function evaluation. An additional procedure will be used in Section~\ref{Sec:Optimal}-C to retrieve the optimal primary ${\bm t}^{\rm opt}$, together with ${\bm Q}^{\rm opt}$.

By combining ${\bm Q}^{*}={\bm 0}$ and Lemmas~\ref{lem.q}--\ref{lem.t}, the dual function $\Phi({\bm \lambda},\mu,\rho,\theta)$ in $({\cal D}1)$ can be evaluated.

\subsection{Obtaining $({\bm \lambda}^{\rm{opt}},\mu^{\rm{opt}},\rho^{\rm{opt}},\theta^{\rm{opt}})$ to Minimize $\Phi({\bm \lambda},\mu,\rho,\theta)$}

Note that the dual function $\Phi({\bm \lambda},\mu,\rho,\theta)$ is convex but non-differentiable in general. As a result, problem $({\cal D}1)$ can be solved by subgradient based methods such as the ellipsoid method\cite{Boyd_Ellipsoid}. For the objective function in (\ref{eq.dual_prob2}a), the subgradient with respect to $(\mv \lambda, \mu,\rho,\theta)$ is given by
$\big[  -\frac{\zeta_1C_1^3q_1^{*3}}{T^2}-\frac{t^*_1}{{\tilde g}_1}\beta\left(\frac{\ell^*_1}{t^*_1}\right)-p_{c,1}t_1^*, \ldots, -\frac{\zeta_KC_K^3q_K^{*3}}{T^2}-\frac{t^*_K}{{\tilde g}_K}\beta\left(\frac{\ell^*_K}{t^*_K}\right)-p_{c,K}t_K^*, T-\sum_{i=1}^K t^*_i,  TP_{\max},L_{\max}-\sum_{i=1}^K \ell_i^*
\big]^\dagger$. As in \cite{Feng17}, the subgradient for the constraint in (\ref{eq.dual_prob2}b) is given by
$
\left[ \eta\mv v^H\mv H_1\mv v,\ldots,\eta\mv v^H\mv H_K\mv v,\;0,\;-1,\;0
\right]^\dagger
$,
where $\mv v\in\mathbb{C}^{N\times1}$ is an eigenvector corresponding to the largest eigenvalue of $\mv G(\mv \lambda,\rho)$ and is given by $\mv v=\argmax_{\|\mv\xi\|=1}\mv\xi^H\mv G(\mv \lambda,\rho)\mv \xi$. Furthermore, the subgradients for the constraints in (\ref{eq.dual_prob2}c) are given by $\mv e_{K+1}$, $\mv e_{K+2}$, $\mv e_{K+3}$, and $\mv e_{i}$, $\forall i\in{\cal K}$, respectively, where $\mv e_{i}$ is the standard unit vector with one in the $i$-th entry and zeros elsewhere in $\mathbb{R}^{(K+3)\times 1}$.

\subsection{Finding $(\boldsymbol{Q}^{\rm{opt}},\boldsymbol{t}^{\rm{opt}},\boldsymbol{\ell}^{\rm{opt}},\boldsymbol{q}^{\rm{opt}})$  to Problem $({\cal P}1)$}

With the optimal dual $(\boldsymbol{\lambda}^{\rm{opt}},\mu^{\rm{opt}},\rho^{\rm{opt}},\theta^{\rm{opt}})$, it remains to determine the optimal solution for $({\cal P}1)$. Specifically, by replacing ${\bm \lambda}$ with ${\bm \lambda}^{\rm{opt}}$ in Lemma~\ref{lem.q}, one can obtain the optimal ${\bm q}^{\rm{opt}}$ to $({\cal P}1)$, and accordingly find the optimal CPU frequencies as $f_{i,1}^{\rm{opt}}=\cdots=f_{i,C_iq_i^{\rm opt}}^{\rm{opt}}=C_iq_i^{\rm{opt}}/T$,~$\forall i\in{\cal K}$. By replacing $(\mv \lambda,\mu,\theta)$ with $(\mv \lambda^{\rm{opt}},\mu^{\rm opt},\theta^{\rm{opt}})$ in Lemma~\ref{lem.t}, we obtain the offloading rate $r_i^{\rm{opt}}$ and accordingly $\ell_i^{\rm opt}=r_i^{\rm opt}t_i^{\rm opt}$, $\forall i\in{\cal K}$. Nevertheless, as $t_i^{*}$ is generally non-unique in (\ref{eq.non-unique-t}), one cannot obtain $t_i^{\rm opt}$ (and $\ell_i^{\rm opt}$) directly here but requires an additional procedure. By substituting ${\bm q}^{\rm opt}$ and $\ell_i=r_i^{\rm opt}t_i$, $\forall i\in{\cal K}$, in problem $({\cal P}1)$, we have the following semidefinite program (SDP) to obtain ${\bm Q}^{\rm opt}$ and ${\bm t}^{\rm opt}$:
\begin{align}\label{eq.prob_Q2}
&(\boldsymbol{Q}^{\rm{opt}},\boldsymbol{t}^{\rm{opt}}) = \argmax_{\boldsymbol{Q}\succeq \boldsymbol{0},\;\boldsymbol{t}}~ \sum_{i=1}^K \omega_i(q_i^{\rm{opt}}+t_ir_i^{\rm{opt}}) \\
 &{\rm s.t.} ~\frac{\zeta_iC_i^3 (q_i^{\rm{opt}})^3}{T^2}+\frac{t_i}{\tilde{g}_i}\beta\left( r_i^{\rm{opt}}\right) + p_{{\rm c},i}t_i \leq T\eta{\rm tr}\left( \boldsymbol{Q}\boldsymbol{H}_i\right),\forall i\in{\cal K} \notag \\
 & ~~~~~\eqref{eqn:Q},~\eqref{eqn:T},~\sum_{i=1}^K t_ir_i^{\rm{opt}} \leq L_{\max},~{\rm and} ~0\leq t_i\leq T,~\forall i\in{\cal K}.\notag
\end{align}
Note that the SDP in \eqref{eq.prob_Q2} can be efficiently solved via CVX \cite{cvx}. Then the optimal $\ell_i^{\rm{opt}}$'s are obtained as $\ell_i^{\rm opt} = r_i^{\rm{opt}}t_i^{\rm{opt}}$, $\forall i\in{\cal K}$. By combining $\mv Q^{\rm opt}$, $\mv t^{\rm opt}$ and $\mv \ell^{\rm opt}$ here, together with $\mv q^{\rm opt}$, the optimal solution to problem $({\cal P}1)$ is found.



\begin{remark}\label{prop2}
It can be readily shown that the optimal solution to $({\cal P}1)$ has the following properties:
\begin{enumerate}
\item For all users, it is optimal to leave a strictly positive number of bits for local computing, i.e., $q_i^{\rm opt}>0$, $\forall i$.
\item At the optimality, each user fully exploits its harvested energy, i.e., the energy harvesting constraints in \eqref{eqn:EH} are active for all users.
\end{enumerate}
 Note that the first property is due to the fact that the marginal energy consumption of local computing is nearly zero when $q_i^{\rm opt}\to 0$; hence, it is always beneficial to leave some bits for local computing. Intuitively, the second property indicates that all users' harvested energy must be used up for $({\cal P}1)$ to maximize the weighted sum-number of computed bits subject to the limited resource and computation latency constraints.
\end{remark}

\section{Numerical Results}\label{Sec:Numerical}
 In this section, numerical results are provided to validate the performance of the proposed design with joint WPT and computation offloading optimization, as compared to the following three benchmark schemes.
 \subsubsection{Local computing only}
 Each user $i\in{\cal K}$ accomplishes its computation task by only local computing. This scheme corresponds to solving $({\cal P}1)$ by setting $\ell_i=0$, $\forall i\in{\cal K}$.
 \subsubsection{Computation offloading only}
 Each user $i\in{\cal K}$ accomplishes its computation task by fully offloading the computation bits to the AP. This scheme corresponds to solving $({\cal P}1)$ by setting $q_i=0$, $\forall i\in{\cal K}$.
 \subsubsection{Joint design with isotropic WPT}
 The AP radiates the RF energy isotropically over all directions by setting $\mv Q=p\mv I$, where $p$ denotes the transmit power at each antenna and $\mv I$ is an $N\times N$ identity matrix. This scheme corresponds to solving $({\cal P}1)$ by replacing $\mv Q$ as $p\mv I$ with $p$ being an optimized variable.

 In the simulations, the AP is equipped with $N=4$ antennas. The energy conversion efficiency is $\eta=0.8$. All channels are modeled as independent Rayleigh fading with an average power loss of $5\times 10^{-6}$ (i.e., $-53$ dB) which corresponds to a distance of about 5 meters from users to the AP in the urban environment. We set $C_i=10^3$ cycles/bit, $\zeta_i=10^{-28}$, and the circuit power as $p_{c,i}=10^{-4}$ Watt (W) for $i\in{\cal K}$. The receiver noise power at the AP is set as $\sigma^2=10^{-9}$~W and the bandwidth for offloading as $B=2$ MHz. We set the maximum number of computation bits supported by the MEC and the maximum CPU frequency for each user $i\in{\cal K}$ as $L_{\max}=2\times 10^5$ bits and $f_{i}^{\max}=0.1$~GHz, respectively. In addition, the weights are set to be identical for different users, i.e., $\omega_i=1/K$, $i\in{\cal K}$. Accordingly, the objective is to maximize the average number of computation bits per user. The results are obtained by averaging over 500 randomized channel realizations. 


\begin{figure}
  \centering
      \vspace{-0.4cm}
  \includegraphics[width = 3.3in]{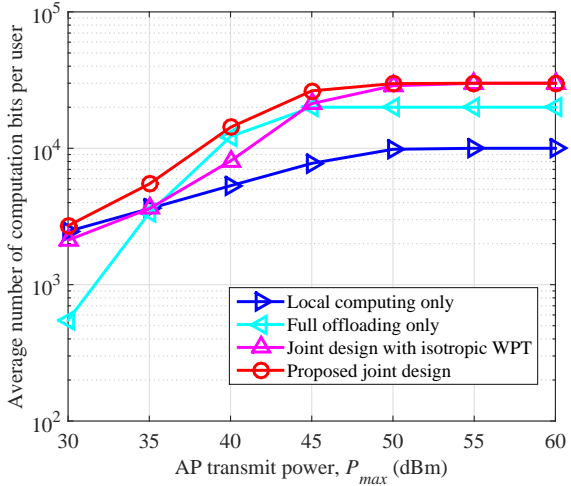}\\
       \vspace{-0.2cm}
  \caption{The average number of computation bits per user versus the maximum transmit power $P_{\max}$ at the AP.}
      \vspace{-0.2cm}    \label{fig.bits_vs_Power}
\end{figure}

\begin{figure}
  \centering
   \vspace{-0.4cm}
  \includegraphics[width = 3.3in]{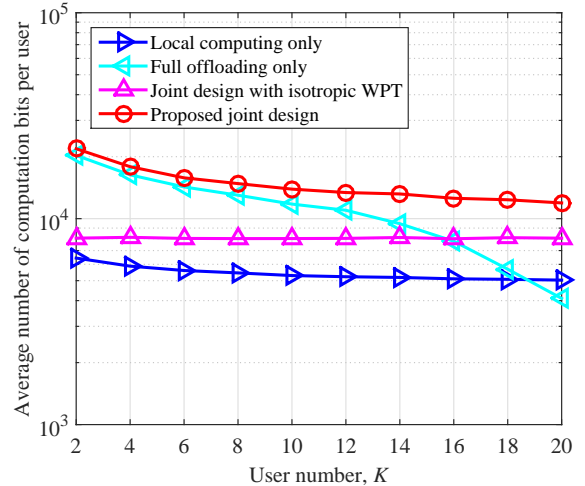}\\
       \vspace{-0.2cm}
  \caption{The average number of computation bits per user versus the user number $K$.}       \vspace{-0.2cm}\label{fig.bits_vs_K}
\end{figure}


 Fig.~\ref{fig.bits_vs_Power} shows the average number of computation bits versus the AP transmit power $P_{\max}$, where $K=10$ and $T=0.1$~sec. It is observed that the proposed joint design achieves significant performance gains over the three benchmark schemes, and the isotropic WPT design is clearly suboptimal. The performances achieved by all the schemes improve significantly as $P_{\max}$ increases. When $P_{\max}\geq 50$~dBm, the average numbers of computation bits for all the schemes are bounded from above. This is expected since in this case, the number of computable bits is fundamentally limited by both the computation capability and users' local CPU frequencies. It is also observed that the local-computing-only scheme achieves a near optimal performance close to that with the proposed joint design at small $P_{\max}$ values. This suggests that most users prefer computing locally in this case.


 Fig.~\ref{fig.bits_vs_K} shows the average number of computation bits versus the user number $K$, where $P_{\max}=40$~dBm and $T=0.1$~sec. In general, we have similar observations as in Fig.~\ref{fig.bits_vs_Power}. Particularly, as $K$ increases, the average number of computation bits per user by all the schemes decreases and the decreasing with the full-offloading-only scheme is more significantly than the other three schemes. This is due to the fact that all users share the finite time block in the full-offloading-only scheme, thereby leading to drastically decreasing of the number of offloaded bits when $K$ becomes large.

\section{Conclusion}\label{Sec:Conclusion}
In this paper, we investigated a unified WPT-MEC design for a wireless powered multiuser MEC system. Specifically, we developed an efficient design framework to maximize the weighted sum of the computation rates across all the users subject to the limited computation/communication resource constraints. Using the Lagrange dual method, we obtained the optimal solution in a semi-closed form. Numerical results demonstrate the merit of the proposed joint design over alternative benchmark schemes. 


 \ifCLASSOPTIONcaptionsoff
  \newpage
\fi

\section*{Appendices}

\subsection{Proof of ${\mv G}({\mv \lambda},\rho)\preceq {\mv 0}$}

The condition ${\mv G}({\mv \lambda},\rho)\preceq {\mv 0}$ can be verified by contradiction. Assume that ${\mv G}({\mv \lambda},\rho)$ is not negative semidefinite. Denote by $\mv \vartheta\in{\mathbb C}^{N\times1}$ an eigenvector corresponding to one positive eigenvalue of ${\mv G}({\mv \lambda},\rho)$. By setting $\mv Q=\tau\mv \vartheta{\mv \vartheta}^H\succeq \mv0$ with $\tau$ going to positive infinity, it follows that
\begin{align}
\lim_{\tau\to +\infty} {\rm tr}({\mv G}({\mv \lambda},\rho)\mv Q)=\lim_{\tau\to+\infty}\tau{\mv \vartheta}^H{\mv G}({\mv \lambda},\rho)\mv\vartheta =+\infty,
\end{align}
which in turn implies that the value $\Phi({\bm \lambda},\mu,\rho,\theta)$ in \eqref{eq.dual_obj2} is unbounded from above over $\mv Q\succeq \mv 0$. Hence, to ensure that $\Phi({\bm \lambda},\mu,\rho,\theta)$ is bounded from above, it requires ${\mv G}({\mv \lambda},\rho)\preceq {\mv 0}$.

\subsection{Proof of Lemma \ref{lem.q}}
Given ${\bm \lambda}$, we solve problem (\ref{eq.La2_q}) for each user $i\in{\cal K}$. When $\lambda_i=0$, the objective function in (\ref{eq.La2_q}) becomes $\omega_iq_i$. It is evident that $q_i^{*}=Tf_i^{\max}/C_i$ is optimal for problem (\ref{eq.La2_q}).

For $\lambda_i>0$, the Lagrangian of (\ref{eq.La2_q}), denoted by $\bar{{\cal L}}_i$, is then given as
\begin{equation}
\bar{{\cal L}}_i = \omega_iq_i-\frac{\lambda_i\zeta_iC_i^3q_i^3}{T^2}+\underline{\eta}_iq_i+\bar{\eta}_i\left(\frac{Tf_i^{\max}}{C_i}-q_i\right),
\end{equation}
where $\underline{\eta}_i\geq 0$ and $\bar{\eta}_i\geq 0$ are the Lagrange multipliers associated with $q_i\geq 0$ and $q_i\leq {Tf_i^{\max}}/{C_i}$, respectively. It can be verified that (\ref{eq.La2_q}) satisfies the Slater's condition. Based on the KKT conditions \cite{BoydBook}, it follows that
\begin{subequations}\label{lem.q.kkt}
\begin{align}
& \underline{\eta}_i^{*} q_i^{*} = 0,~~\bar{\eta}^{*}_i\left(\frac{Tf_i^{\max}}{C_i}-q^{*}_i\right)=0\\
& \omega_i-\frac{3\lambda_i\zeta_iC_i^3(q_i^{*})^2}{T^2}+\underline{\eta}_i^{*}-\bar{\eta}^{*}_i = 0,
\end{align}
\end{subequations}
where $\underline{\eta}_i^{*}$ and $\bar{\eta}_i^{*}$ are the optimal dual variables, (\ref{lem.q.kkt}a) collects the complementary slackness conditions, and (\ref{lem.q.kkt}b) is the first-order derivative condition for $\bar{{\cal L}}_i$ with respect to $q_i$. From (\ref{lem.q.kkt}a) and (\ref{lem.q.kkt}b), it thus follows that
\begin{equation}
q_i^{*} =  \min \left( \sqrt{\frac{\omega_iT^2}{3\lambda_i\zeta_iC_i^3}},\; \frac{Tf_i^{\max}}{C_i} \right).
\end{equation}

\subsection{Proof of Lemma \ref{lem.t}}
When $\lambda_i=0$, the objective function in (\ref{eq.La2_t_l}) becomes $(\omega_i-\theta)\ell_i-\mu t_i$. Evidently, if $\omega_i-\theta >0$, it follows that $t_i^{*}=0$ and $\ell_i^{*}=L_{\max}$ are optimal for (\ref{eq.La2_t_l}); otherwise, $t_i^{*}=\ell_i^{*}=0$.

For $\lambda_i>0$, the Lagrangian of (\ref{eq.La2_t_l}) is given by
\begin{align}\label{eq.La_t_l}
 {\underline {\cal L}}_i =& (\omega_i-\theta)\ell_i - \frac{\lambda_it_i}{\tilde{g}_i} \beta\left(\frac{\ell_i}{t_i}\right) - \mu t_i - \lambda_ip_{{\rm c},i}t_i + \bar{a}_it_i \notag \\
 &+\bar{b}_i \ell_i + \bar{c}_i (T-t_i)
\end{align}
where $\bar{a}_i\geq 0$, $\bar{b}_i\geq 0$, and $\bar{c}_i\geq 0$ are the Lagrangian multipliers associated with $t_i\geq 0$, $\ell_i\geq 0$, and $t_i\leq T$, respectively.

Based on the KKT conditions\cite{BoydBook}, it follows that
\begin{subequations} \label{eq.t_l_kkt}
\begin{align}
& \bar{a}_i^{*}t_i^{*}=0,~~\bar{b}^{*}_i\ell_i^{*}= 0,~~\bar{c}_i^{*}(T-t_i^{*})=0 \\
&  -\frac{\lambda_i}{\tilde{g}_i}\left( \beta\left(\frac{\ell_i^{*}}{t_i^{*}}\right) - \frac{\ell_i^{*}}{t_i^{*}}\beta' \left(\frac{\ell_i^{*}}{t_i^{*}}\right) \right) - \mu -\lambda_ip_{{\rm c},i} + \bar{a}_i^{*} -\bar{c}_i^{*}=0 \\
& (\omega_i-\theta) - \frac{\lambda_i}{\tilde{g}_i} \beta'\left(\frac{\ell_i^{*}}{t_i^{*}}\right)+\bar{b}_i^{*}=0,
\end{align}
\end{subequations}
where $(\bar{a}_i^{*},\bar{b}_i^{*},\bar{c}_i^{*})$ denotes the optimal dual solution, (\ref{eq.t_l_kkt}a) collects the complementary slackness conditions, and the left-hand-side (LHS) terms of (\ref{eq.t_l_kkt}b) and (\ref{eq.t_l_kkt}c) are the first-order derivatives of ${\underline {\cal L}}_i$ with respect to $t_i^{*}$ and $\ell_i^{*}$, respectively. Let $r_i^{*} \triangleq {\ell_i^{*}}/{t_i^{*}}$, and define $r_i^{*}=0$ if either $\ell_i^{*}=0$ or $t_i^{*}=0$. From (\ref{eq.t_l_kkt}a) and (\ref{eq.t_l_kkt}c), we have
\begin{equation}\label{eq.r_i_kkt}
r_i^{*} = \left\{ \begin{aligned}
&0,&&{\rm if}~ (\omega_i-\theta)B\tilde{g}_i\leq \lambda_i\sigma^2\ln2 \\
& B\log_2\left( \frac{(\omega_i-\theta)B\tilde{g}_i}{\lambda_i\sigma^2\ln 2}
\right), &&\rm{otherwise}.\\
\end{aligned} \right.
\end{equation}
Furthermore, substituting (\ref{eq.r_i_kkt}) into (\ref{eq.t_l_kkt}b), we obtain that
\begin{equation}
\bar{a}_i^{*} -\bar{c}_i^{*} = \frac{\lambda_i}{\tilde{g}_i}\left( \beta\left(r_i^{*}\right) - r_i\beta' \left(r_i^{*}\right) \right) + \mu + \lambda_ip_{{\rm c},i}.
\end{equation}
Clearly, $\bar{a}_i^{*} -\bar{c}_i^{*}>0$ and $\bar{a}_i^{*} -\bar{c}_i^{*}< 0$ imply that $\bar{a}_i^{*}>0$ and $\bar{c}_i^{*}> 0$, respectively. In addition, when $\bar{a}_i^{*} -\bar{c}_i^{*}=0$, we have $\bar{a}_i^{*} = \bar{c}_i^{*}=0$. Based on the complementary slackness conditions in (\ref{eq.t_l_kkt}a), it follows that
\begin{equation*}\label{eq.t_i**}
t_i^{*}
\left\{ \begin{aligned}
&=0,&&{\rm if} ~ \frac{\lambda_i}{\tilde{g}_i}\left( \beta\left(r_i^{*}\right) - r_i\beta' \left(r_i^{*}\right) \right) + \mu + \lambda_ip_{{\rm c},i} > 0 \\
&\in[0,T],&&{\rm if} ~ \frac{\lambda_i}{\tilde{g}_i}\left( \beta\left(r_i^{*}\right) - r_i\beta' \left(r_i^{*}\right) \right) + \mu + \lambda_ip_{{\rm c},i} = 0 \\
&=T,&&{\rm otherwise},
\end{aligned} \right.
\end{equation*}
and $\ell_i^{*}=r_i^{*}t_i^{*}$. Now it completes the proof of Lemma~\ref{lem.t}.


 \end{document}